\documentclass[fleqn,usenatbib]{mnras}

\usepackage{newtxtext,newtxmath}
\usepackage[T1]{fontenc}

\DeclareRobustCommand{\VAN}[3]{#2}
\let\VANthebibliography\thebibliography
\def\thebibliography{\DeclareRobustCommand{\VAN}[3]{##3}\VANthebibliography}

\usepackage{graphicx}	% Including figure files
\usepackage{amsmath}	% Advanced maths commands

\newcommand{\fermi}{{\it Fermi}-LAT}
\newcommand{\gray}{$\gamma$-ray}
\newcommand{\grays}{$\gamma$-rays}
\newcommand{\fgl}{4FGL J1544.3-0649}

\title[The strange case of 4FGL J1544.3-0649]{The strange case of the transient HBL blazar 4FGL  J1544.3-0649}

\author[N. Sahakyan and P. Giommi]{
N. Sahakyan,$^{1, 2}$\thanks{E-mail: narek@icra.it},
P. Giommi,$^{2, 3, 4,5}$\thanks{E-mail: giommipaolo@gmail.com}\\
$^{1}$ICRANet-Armenia, Marshall Baghramian Avenue 24a, Yerevan 0019, Armenia\\
$^{2}$ICRANet, P.zza della Repubblica 10, 65122 Pescara, Italy\\
$^{3}$ Institute for Advanced Study, Technische Universit{\"a}t M{\"u}nchen, Lichtenbergstrasse 2a, D-85748 Garching bei M\"unchen, Germany\\
$^{4}$ Excellence Cluster ORIGINS, Boltzmannstrasse 2, D-85748 Garching bei M\"unchen, Germany\\
$5$ Associated to Italian Space Agency, ASI, via del Politecnico snc, 00133 Roma, Italy
}

\date{Accepted XXX. Received YYY; in original form ZZZ}

\pubyear{2020}

\begin{document}
\label{firstpage}
\pagerange{\pageref{firstpage}--\pageref{lastpage}}
\maketitle

\begin{abstract}
We present a multifrequency study of the transient \gray\ source \fgl, a blazar that exhibited a remarkable behaviour raising from the state of an anonymous mid-intensity radio source, never detected at high energies, to that of one of the brightest extreme blazars in the X-ray and \gray\ sky. Our analysis shows that the averaged $\gamma$-ray spectrum is well described by a powerlaw with a photon index of $1.87\pm0.04$, while the flux above 100 MeV is $(8.0\pm0.9)\times10^{-9}\:{\rm photon\: cm^{-2}\: s^{-1}}$, which increases during the active state of the source. The X-ray flux and spectral slope are both highly variable, with the highest 2-10 keV flux reaching $(1.28\pm0.05)\times10^{-10}\:{\rm erg\:cm^{-2}\:s^{-1}}$. On several observations the X-ray spectrum hardened to the  point implying as SED peak moving to energies larger than 10 keV. As in many extreme blazars the broadband spectral energy distribution can be described by a homogeneous one-zone synchrotron-self-Compton leptonic model. 
We briefly discuss the potential implications for high-energy multi-messenger astrophysics in case the dual behaviour shown by \fgl\ does not represent an isolated case, but rather a manifestation of a so far unnoticed relatively common phenomenon.
\end{abstract}


%%%%%%%%%%%%%%%%%%%%%%%%%%%%%%%%%%%%%%%%%%%%%%%%%%

%%%%%%%%%%%%%%%%% BODY OF PAPER %%%%%%%%%%%%%%%%%%

\section{Introduction}
%{\color{blue} The text in blue should be still corrected.}\\
Blazars are a special type of active galactic nuclei (AGN) hosting a relativistic jet that happens to be pointing very close to the observer's line of sight \citep[e.g.][]{Blandford1978,AGNReview}. The emission in these sources is dominated by non-thermal radiation across the electromagnetic spectrum, often exhibiting large variations on timescales ranging from years to a few hundred seconds. Blazars are usually grouped into BL Lacertae (BL Lac) objects and Flat Spectrum Radio Quasars (FSRQs), based on their properties in the optical band: the optical spectrum of FSRQs shows broad emission lines while that of BL Lacs is  featureless, shows narrow lines or features of the host galaxy \citep{UrryPadovani}.\\
Blazars are the most luminous and energetic persistent emitters in the known Universe, dominating in the extragalactic \gray\ sky.
Among the total $\sim5,800$ sources in the Fermi Large Area Telescope (\fermi) fourth source catalog  of \gray\ sources \citep[4FGL-DR2 ;][]{2020ApJS..247...33A,4LAC,4FGL-DR2,4LAC-DR2} $\sim3,100$ are blazars, 47 are radio galaxies, and 19 are other types of AGNs. Interestingly, \gray\ emission has been detected also from blazars at very high redshifts \citep[e.g.,][]{2017ApJ...837L...5A, 2020MNRAS.498.2594S} which enables the study of the evolution of the most luminous relativistic jets over cosmic time.\\
The broadband spectral energy distribution (SED) of blazars has a characteristic double hump shape. The low energy peak (radio to UV/X-ray) is usually explained by synchrotron emission of relativistic electrons propagating along the jet, although recent models developed to explain the emission of high-energy neutrinos in blazars assume that this component may be due to proton synchrotron radiation, at least during flares \citep{MP}.
The second SED component extends to High Energies (HEs; $>100$ MeV), its origin being debatable. In the leptonic scenarios, this component is interpreted as inverse Compton upscattering of synchrotron photons (Synchrotron self Compton [SSC] \citet{1992ApJ...397L...5M, bloom, ghisellini}) or of external photons (external inverse Compton) \citep{sikora,1994ApJS...90..945D}. On the other hand, in hadronic models the HE component is mainly due to proton synchrotron emission \citep{Aharonian2000,2001APh....15..121M} or pion decay \citep{1993A&A...269...67M, 1989A&A...221..211M, 2001APh....15..121M, mucke2, 2013ApJ...768...54B}. In the latter case, blazars are also sources of Very High Energy (VHE; $>100$ GeV) neutrinos \citep{2018Sci...361..147I, 2018Sci...361.1378I, 2018ApJ...863L..10A,2018ApJ...864...84K, 2018ApJ...865..124M, 2018MNRAS.480..192P, 2018ApJ...866..109S, 2019MNRAS.484.2067R,2019MNRAS.483L..12C,2019NatAs...3...88G}.\\
Depending on the peak of the low energy component ($\nu_{\rm peak-s}$) blazars can be further grouped into three subcategories, namely low-energy peaked blazars (LBLs; $\nu_{\rm peak-s}<10^{14}$ Hz), intermediate peaked blazars (IBLs; $10^{14}\:{\rm Hz}<\nu_{\rm peak-s}<10^{15}$ Hz) and high-energy peaked blazars (HBLs; $\nu_{\rm peak-s}>10^{15}$ Hz) \citep{Padovani1995,Abdo_2010}. Observations in the X-ray band reveal that sometimes the synchrotron peak of HBLs can reach energies of $\sim$1 keV, ($\sim 2\times 10^{17}$ Hz) or even larger, showing an extreme behaviour \citep[extreme HBLs, e.g.][]{sedentary,2001A&A...371..512C,Biteau2000}. For example, during the flares of Mkn 501 the synchrotron peak reached $\sim 100$ keV \citep{1998ApJ...492L..17P} or during the flares of 1ES 1218+304 the X-ray spectral index hardened to $\Gamma \leq1.80$ shifting the peak towards higher energies \citep{2020MNRAS.496.5518S}.
The third catalog of high energy peaked blazars \citep[3HSP,][]{3HSP} includes several objects in this category.
Recent observations in the HE and VHE \gray\ bands have revealed an additional class of BL Lac objects VHE \gray\ spectrum of which is characterized by a hard intrinsic photon index of up to 1 TeV after correction for the extragalactic background light (EBL) absorption effects (BL Lacs extreme in \grays;  \citet{2015MNRAS.451..611B,2011MNRAS.414.3566T}).\\
The BL Lacs having featureless optical spectra have been traditionally discovered in radio or X-ray surveys \citep[e.g.,][]{2005A&A...434..385G}. Currently, the \gray\ data from \fermi\ observations combined with the data at the lower energy bands  \citep[e.g., infrared, IR, ][]{2012ApJ...752...61M} are a powerful additional method; nearly 38\% of detected blazars are BL Lacs in the fourth catalog of AGNs detected by \fermi\ \citep{3HSP,2020ApJ...892..105A}. Nearly all these \gray\ emitting BL Lacs have detected radio counterparts, which makes radio observations crucial also for the understanding of their physics. In this regard of a particular interest is the identification of radio-weak BL Lac objects, i.e., BL Lacs from which radio emission is detected in follow-up observations but which do not have counterparts in the major radio surveys. A handful observations of radio-weak BL Lac objects will have an impact on our understanding of AGN unification as well as on BL Lac associations based on the radio surveys \citep{2017ApJ...834..113M,3HSP}.\\
%\textcolor{red}{@Paolo: maybe we could add more here on BL Lacs ?}.\\
In May 2017 \fermi\ detected a new \gray\ source (\fgl) not associated with any previously known \gray\ object; it was bright in the \gray\ band in two consecutive weeks after the discovery \citep{2017ATel10482....1C}. The X-ray follow-up observations by Neil Gehrels Swift Observatory \citep{2004ApJ...611.1005G}, (hereafter Swift) found a new bright X-ray source at a position corresponding to the optical transient ASASSN-17gs detected at $V=17.3$ mag on May 25 \citep{2017ATel10482....1C}. The MAXI team \citep{MAXI} also reported a detection of this source on May 12-13 at a flux level similar to that observed by Swift, with a flux increase of a factor $\sim$2 in the period May 21 to May 25.
The spectroscopic redshift of the host galaxy, 2MASX J15441967-0649156, was estimated to be $z=0.171$ using the MDM 2.4m Hiltner telescope \citep{2017ATel10491....1C}. The source has a counterpart in the radio band with flux densities of 46.6 mJy at $1.4$ GHz \citep[NRAO VLA Sky Survey;][]{1998AJ....115.1693C}, 35.8 mJy at 3 GHz \citep[Very Large Array Sky Survey, epoch 1;][]{VLASS1} and 67 mJy at 150 MHz \citep[TIFR GMRT Sky Survey;][]{ 2017A&A...598A..78I}. In  addition, \citet{2018ApJ...854L..23B} monitored \fgl\ with the Effelsberg-100 m single dish radio telescope for a four-month follow-up after the brightening in the \gray\ band.
%This showed a spectral index of $>-0.5$ ($S=\nu^{\alpha}$) between 4.85 and 14.60 GHz suggesting a blazar-like emission with the jet oriented toward the observer. 
These data imply a flat radio spectrum very similar to that of most blazars, and a radio luminosity of $\sim 5\times 10^{40}$ erg s$^{-1}$, at 1.4 GHz, more than a factor 10 larger than that of Mrk 501 and Mrk 421.
Considering the post-burst data at 5 GHz, the radio to X-ray flux ratio is log (F$_{\rm 1.4\:GHz}$/F$_{\rm 1\:kev}$) = -4.3 
%\textcolor{green}{Give units: I guess it is log (F$_{\rm 1.4\:GHz}$/F$_{\rm 1\:kev}$) = -4.3?}
%at the border between the radio-loud and radio-quiet AGN populations 
typical of HBL Blazars. 
Follow up XMM-Newtown (XMM) observations showed that the source exhibits strong variability both in the X-ray flux and spectral shape on timescales of $\sim10$ ks and the X-ray spectrum is described by a broken powerlaw \citep{2019A&A...622A.116U}.\\
%\textcolor{red}{@Paolo: what else we should add here ?}.\\
The transient 
%and radio-week 
nature of \fgl\ makes it a unique object and an interesting target for broadband studies. Understanding of the changes in the jet of \fgl\ which led to its brightening in the X-ray and \gray\ bands can shed light on the physics at work in the BL Lac jets. 
 %\textcolor{green}{On the other hand, if \fgl belongs to the class of radio-weak BL Lacs which are challenging the current unification models of AGNs, its detailed multiwavelength spectral and temporal studies can be crucial for understanding of AGNs in general.
 %\bf{PG this source a ten time brighter than MRK 501, it is not radio weak... I would remove this phrase}}. 
 In this regard, the many monitoring observations of \fgl\ with Swift provide unprecedented data both in the optical/UV and X-ray bands while that of \fermi\ in the 100 MeV-500 GeV band allowing to shape the low and high energy components in the SED of \fgl\ and investigate its evolution in time. Also, this broadband data can be used for theoretical modeling allowing to derive the main parameters characterizing the jet of \fgl.\\
The purpose of this paper is to investigate the origin of the  broadband emission from \fgl\ and the astrophysical multi-messenger implications of the dual behaviour of this source. % by analyzing the data in the optical/UV, X-ray and \gray bands. 
The paper is organized as follows. The \fermi\ and Swift data extraction and analyses are described in Section \ref{anal}. In Section \ref{mwemiss} the origin of the multiwavelength emission is investigated, and the discussion and conclusion are given in Section \ref{diss}.

\section{Data analyses}\label{anal}
\subsection{\fermi\ observations of \fgl}
The \gray\ data used in this work was collected by \fermi\ from August 2008 to July 2020 (from MJD 54683 to MJD 59058). The Pass 8 SOURCE data in the energy range from 100 MeV to 600 GeV have been downloaded and analyzed using the standard reduction methodology suggested by the LAT collaboration. The events classified as {\it evclass=128} and {\it evtype=3} within a circular region of $12^{\circ}$ centered on the \gray\ position of \fgl\ (R.A., decl.=236.078, -6.825) are analyzed using Fermi ScienceTools (1.2.1) with P8R3\_ SOURCE\_ V2 instrument response functions. The good time intervals were selected with the filter "DATA\_ QUAL>0" and "LAT\_ CONFIG==1" in the {\it gtmktime} tool whereas the Earth limb photons were removed by applying a maximum zenith angle cut of $90^{\circ}$. The data are divided into 38 logarithmically equal energy bins and binned into the $16.9^{\circ}\times16.9^{\circ}$ square region of interest. Then, binned likelihood analysis implemented in {\it gtlike} tool was used to estimate the \gray\ flux and photon index of the sources. The model file is generated from the \fermi\ fourth source catalog of \gray\ sources (4FGL) using the 4FGL-DR2 version of the 4FGL which is based on 10 years of survey; it includes all point like sources within $12^{\circ}+5^{\circ}$ around \fgl\ together with the Galactic and extragalactic diffuse \gray\ models ({\it gll\_iem\_v07} and {\it iso\_P8R3\_SOURCE\_V2\_v1}, respectively). The spectral shapes of all sources are adopted from the 4FGL catalog and all spectral parameters of the sources falling between $12^{\circ}$ and $12^{\circ}+5^{\circ}$ are kept fixed. The \gray spectrum of \fgl is modeled with a power-law function. The significance of the \gray\ signal is evaluated by the test statistic ($TS$) \citep{1996ApJ...461..396M} defined as $TS = 2\times(log L_{1} -log L_0)$ where $L_{1}$ and $L_{0}$ are the likelihood of the data with and without a point source at the position of the source under investigation.\\
The \gray\ light curves were calculated by performing an unbinned maximum likelihood analysis with the appropriate quality cuts as described above. Initially the light curve was computed by dividing the entire observation time into 21-day intervals and selecting events in the energy range from 500 MeV to 300 GeV. The lower limit was selected 500 MeV, since due to the hard \gray\ photon index of \fgl\ the source is undetected below $\sim500$ MeV. Also, the normalization of both diffuse components was fixed to the values obtained for the whole time period, as no variability is expected from the background. Next, the \gray\ variability of the \fgl\ was investigated by generating the light curves with the help of an adaptive binning method \citep{2012A&A...544A...6L}. This produces light curves with flexible time bins, requiring a fixed uncertainty for the flux estimation in each time interval. The fluxes are computed above the optimal energy, which is $E_{\rm p}=801.6$ MeV in this case. This allows a detailed investigation of the \gray\ flux variation and is a powerful method for identifying flaring periods (if any).\\
\begin{figure*}
%   \centering
   \includegraphics[width=0.99 \textwidth]{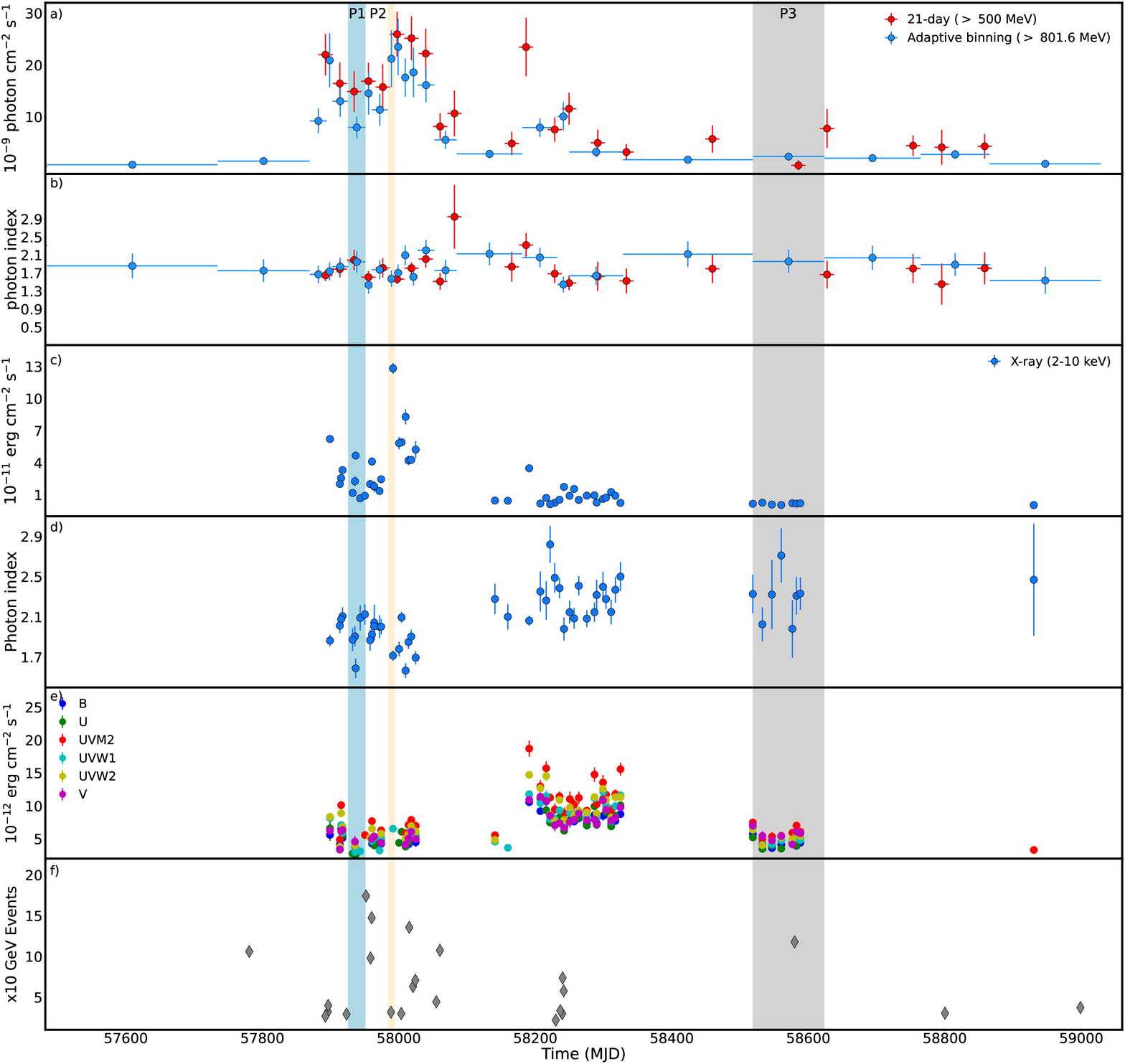}\\

      \caption{Multiwavelength light curves for \fgl.}
         \label{lightcurve}
\end{figure*}
The analysis of entire $\sim11.9$ years data yielded the source detection with ${\rm TS=778.7\:(27.9\:\sigma)}$ and with a hard \gray\ spectrum ($\sim E^{-1.87\pm0.04}$) consistent with the value given in 4FGL. The integral flux estimated above 100 MeV is $(8.0\pm0.9)\times10^{-9}\:{\rm photon\:cm^{-2}\:s^{-1}}$. Panels a) and b) of Fig. \ref{lightcurve} show the \gray\ flux and photon index evolution in time calculated for 21-day (red) bins where only the intervals when $TS>12$ are presented. %The \gray\ light curves show that the source %detection significance is below the threshold of TS$=25$ before MJD 57484 and then it 
The source was in the bright \gray\ flaring state around MJD 58000 with the highest \gray\ flux of $(2.59\pm0.43)\times10^{-8}\:{\rm photon\:cm^{-2}\:s^{-1}}$ observed on MJD $57997.5\pm10.5$ with a detection significance of $15.0\:\sigma$. Yet, the adaptively binned light curve illustrates the changes in the \gray\ flux better; the source being mostly undetected in the \gray\ band shows significant ($>5\sigma$) \gray\ emission with a flux of $(7.75\pm2.96)\times10^{-10}\:{\rm photon\:cm^{-2}\:s^{-1}}$ above $E_{\rm p}=801.6$ MeV starting from MJD $57484.7$. However, the \gray\ flux of the source substantially increased during MJD 57869.3-58052.0; the highest \gray\ flux of $(2.35\pm	0.55)\times10^{-8}\:{\rm photon\:cm^{-2}\:s^{-1}}$ above $E_{\rm p}=801.6$ MeV was observed on MJD $57998.9\pm 4.60$. The flux was above $10^{-8}\:{\rm photon\:cm^{-2}\:s^{-1}}$ in other eight periods (see Fig. \ref{lightcurve}). Despite the increase in the \gray\ flux, the photon index does not vary significantly (panel b in Fig. \ref{lightcurve}); even though there are periods when the spectrum hardens, no definite conclusion can be drawn because of large uncertainties. For example, in the adaptively binned intervals the hardest indexes of $1.43\pm0.19$ and $1.44\pm0.18$ were observed on MJD $57955.64\pm4.49$ and MJD $58241.06\pm8.70$ which within the uncertainty agree with the averaged photon index estimated during $\sim11.9$ years. Interestingly, in the 21-day binned light curve, when the highest \gray\ flux was observed on MJD $57997.5\pm10.5$, the \gray\ spectrum is hard with a photon index of $1.57\pm0.11$.\\
The distribution of HE photons arrival time (computed using the {\it gtsrcprob} tool) is shown in panel f) of Fig. \ref{lightcurve}. The highest energy event with $E_{\gamma}=174.20$ GeV has been observed on MJD 57951.92 within a circle of $0.054^{\circ}$ around \fgl, with the probability of $0.99428$ being associated with it. There are four additional events from the innermost region around \fgl\ with an energy exceeding 100 GeV; the 107.4, 106.0, 147.4 and 117.8 GeV events were observed on MJD 58060.2, 57780.9, 57960.1 and 58579.9, respectively.
\subsection{Swift XRT observations}

\fgl\, was pointed by Swift 51 times, starting shortly after the discovery of the \gray\ and optical transient in May 2017 to mid 2020. All the XRT data was processed using Swift$\_$xrtproc, a data reduction, spectral and imaging analysis tool, which automatically downloads the data from one of the Swift archives, generates high-level calibrated data products, and performs a comprehensive spectral and imaging analysis of a source, using the XSPEC12.11 and XIMAGE4.5.1 packages, released as part of the HEASoft-6.28 distribution.
Swift$\_$xrtproc was developed within the Open Universe initiative \footnote{https://openuniverse.asi.it} \citep{GiommiOU}, in collaboration with the ASI Space Science Data Center (SSDC) \footnote{https://www.ssdc.asi.it}, to analyse all the XRT observations of a large sample of blazars and the data of all the Gamma Ray Bursts (GRBs) observed by Swift-XRT (Giommi et al. 2020 in preparation). 

\begin{figure}
%   \centering
   \includegraphics[width=0.49 \textwidth]{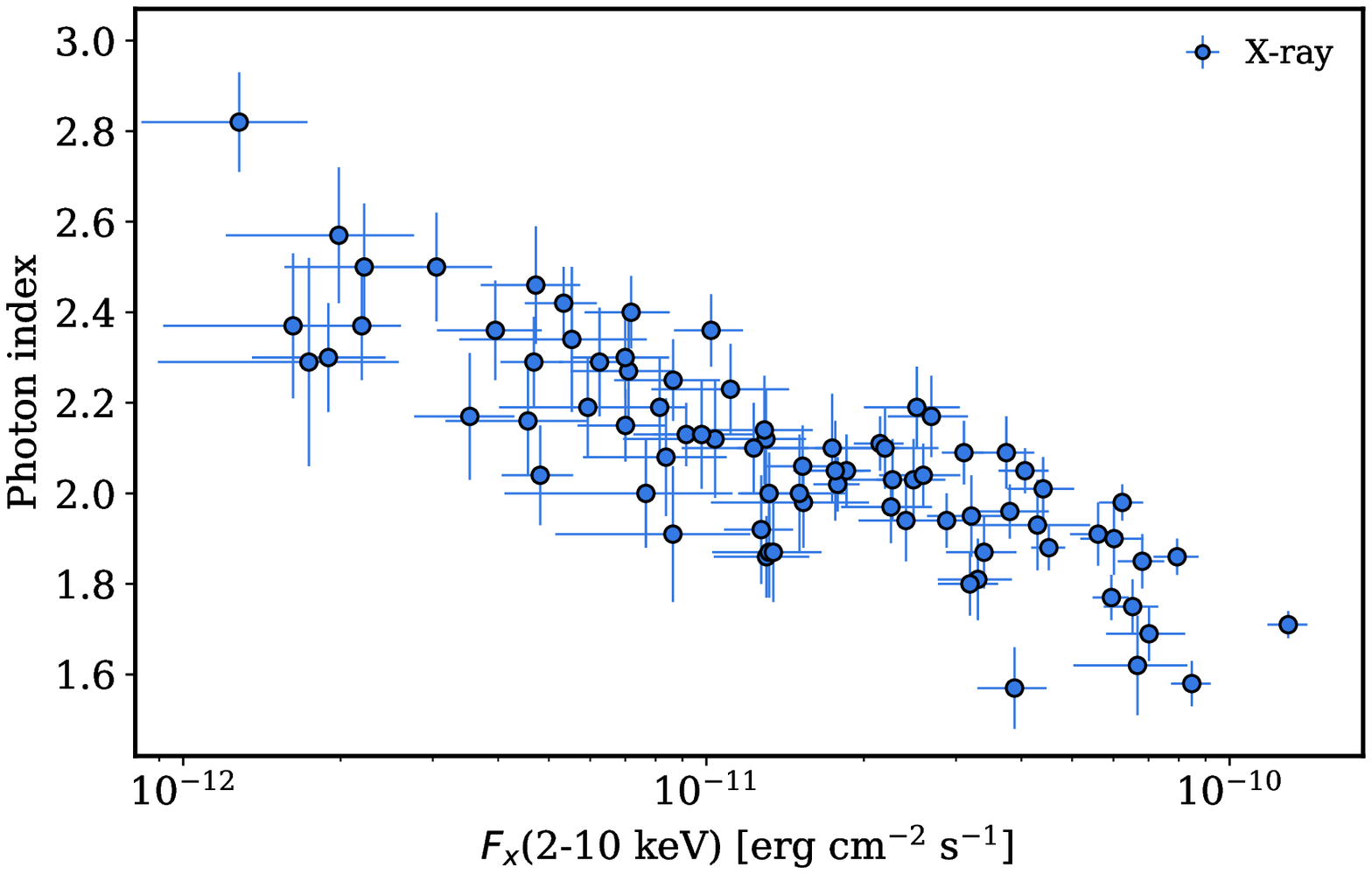}\\

      \caption{The X-ray flux versus photon index.%\textcolor{red}{The label on the X-axis is not photons but erg/cm2/s; The energy band need to be stated (e.g. 2-10 keV). It could be that if we used the log on the X-axis the plot looks even better...}
      }
         \label{Xindex}
\end{figure}

The 2-10 keV X-ray flux and photon index (0.3-10 keV) are shown in Fig. \ref{lightcurve} c) and d), respectively. The source was in a high X-ray state contemporaneously with the \gray\ flare around MJD 58000 with the highest X-ray flux of $(1.28\pm0.05)\times10^{-10}\:{\rm erg\:cm^{-2}\:s^{-1}}$ observed on MJD 57991.40. Then, the source is in a quiescent state with the flux of the order of $(1.0-5.0)\times10^{-12}\:{\rm erg\:cm^{-2}\:s^{-1}}$. The X-ray photon index shows an interesting evolution in time; the X-ray spectrum is hard ($<2.0$) during the bright X-ray state around MJD 58000 which then softens during the moderately bright or quiescent states. The hardest X-ray photon index of $1.57\pm 0.07$ was observed on MJD 58009.92; there are additional 12 periods with an X-ray photon index of $\leq1.9$. This hardening clearly shifts the peak of synchrotron component to higher energies. The relation between the X-ray flux and the photon index was further investigated by considering XRT observations in each orbit. The X-ray flux (2-10 keV) versus the photon index is shown in Fig. \ref{Xindex}. The spectrum clearly tends to harden as the source gets brighter. The linear-Pearson correlation test yields $r_{\rm d}=-0.71$ which suggests a negative correlation between the flux and photon index, i.e., when the photon index decreases (hardens) the flux increases.
\subsection{Swift UVOT}
Along with the XRT observations, the source was also targeted with Swift UVOT in the ultraviolet UVW1, UVM2, and UVW2 and optical V, B and U bands. For each observation, the aperture photometry analysis was performed using the standard UVOT software distributed within the HEASoft package (v6.26.1) and the calibration included in the latest release of CALDB. Source counts were extracted from a circular region with $5''$ radius, and the background ones from a source-free region with $20''$ radius. Each individual observation was checked to validate that in any filter the background was not contaminated by the nearby objects. The counts were extracted and converted to fluxes using UVOTSOURCE tool and the conversion factors from \citet{2008MNRAS.383..627P}. The flux values were then de-reddened using the value of E (B - V ) = 0.138 \citep{2011ApJ...737..103S}. \\
The optical/UV light curve is shown in Fig. \ref{lightcurve} e). The source appears to be in a quiescent state in the optical/UV bands when it was bright in the X-ray and \gray\ bands (MJD 58000). However, the optical/UV flux substantially increased around MJD $\sim58250$. The highest flux of $(1.87\pm0.12)\times10^{-11}\:{\rm erg\:cm^{-2}\:s^{-1}}$ was observed on MJD 58191.08 in UVM2 filter. In this period the flux increased in other UVOT filters as well, being at the level of $(1.06-1.47)\times10^{-11}\:{\rm erg\:cm^{-2}\:s^{-1}}$. 
%Interestingly, during this active optical/UV period coincides with brightening in the X-ray band whereas in the \gray band the flux changed only marginally.
%\subsection{ASAS-SN}
%\subsection{Evolution of Multiwavelength SED}
%The evolution of emission spectra in time are investigated by SED/light curve movie which is an animation on the simultaneous data. The \gray data for each adaptively binned interval is combined with the optical/UV and X-ray data from the Swift observations and with the Effelsberg-100 m observations of \fgl in the frequencies from 5 to 15 GHz \citep{2018ApJ...854L..23B}. Moving over each \gray light curve interval, the spectrum observed in different bands evolves showing the time evolution of the SED. The SED/light curve animation is available here.  Before the large \gray flare on MJD 58000 the source is barely detected in the \gray band, i.e., the SEDs on MJD 57274.73 and  MJD 57274.73 contain upper limits and large uncertainties in the flux estimation. Then the \gray emission substantially increases which is sometimes accompanied by increases in the other bands. The \gray emission from the source is still detectable with a high detection significance until MJD 58865.87, but in the last interval of the considered period the detection significance drops to $TS=17$.
\section{The origin of multiwavelength emission}\label{mwemiss}
Multiwavelength observations of \fgl\,  allow us to build the SEDs with simultaneous and archival data as well as to investigate the time evolution of SEDs. The \gray\ data for each adaptively binned interval has been combined with the Swift optical/UV and X-ray data, with the Effelsberg-100 m observations of \fgl\, \citep[in the frequencies from 5 to 15 GHz ][]{2018ApJ...854L..23B} and with archival data from the VOU-Blazar \citep{VOU-Blazars} and the SSDC\footnote{https://tools.ssdc.asi.it/SED} SED tools, to form the SED/light curve animation that is available here \href{https://www.youtube.com/watch?v=9feaNWi1RDs}{\nolinkurl{youtu.be/9feaNWi1RDs}}. Moving over each \gray\ light curve interval, the spectrum observed in different bands evolves showing the time evolution of the SED. The months before the large \gray\ flare on MJD 58000 (May 2017) the source is barely detected in the \gray\ band, i.e., the SED on MJD 57274.73 contain upper limits and large uncertainties in the flux estimation. Then the \gray\ emission substantially increases which is sometimes accompanied by increases in the other bands. The \gray\ emission from the source is still detectable with a high significance until MJD 58865.87, but in the last interval (MJD $58947.23\pm 81.36$) the detection significance drops to $TS=17$. 
To investigate the nature of the multiwavelength emission from \fgl, data from the following periods are considered:
\begin{itemize}
%\item[]  MJD 57898.7: high \gray emission state on MJD $57898.7\pm4.0$ coinciding with Obsid: 10145001 Swift observation.
\item[] Period 1 (P1): averaged \gray\ emission state on MJD $57938.35\pm12.80$ with contemporaneous Swift UVOT/XRT (Obsid: 10145008) and radio data. 
\item[] Period 2 (P2): from MJD 57984.00 to MJD 57994.33 corresponding to the period when the highest X-ray flux was observed (Obsid: 10145018).
\item[] Period 3 (P3): 105.06-days interval centered on MJD $58571.12$ when the source was in a low state in the \gray, X-ray and optical/UV bands (Obsid: 10145057).

\end{itemize}

\begin{figure*}
%   \centering
   \includegraphics[width=0.49 \textwidth]{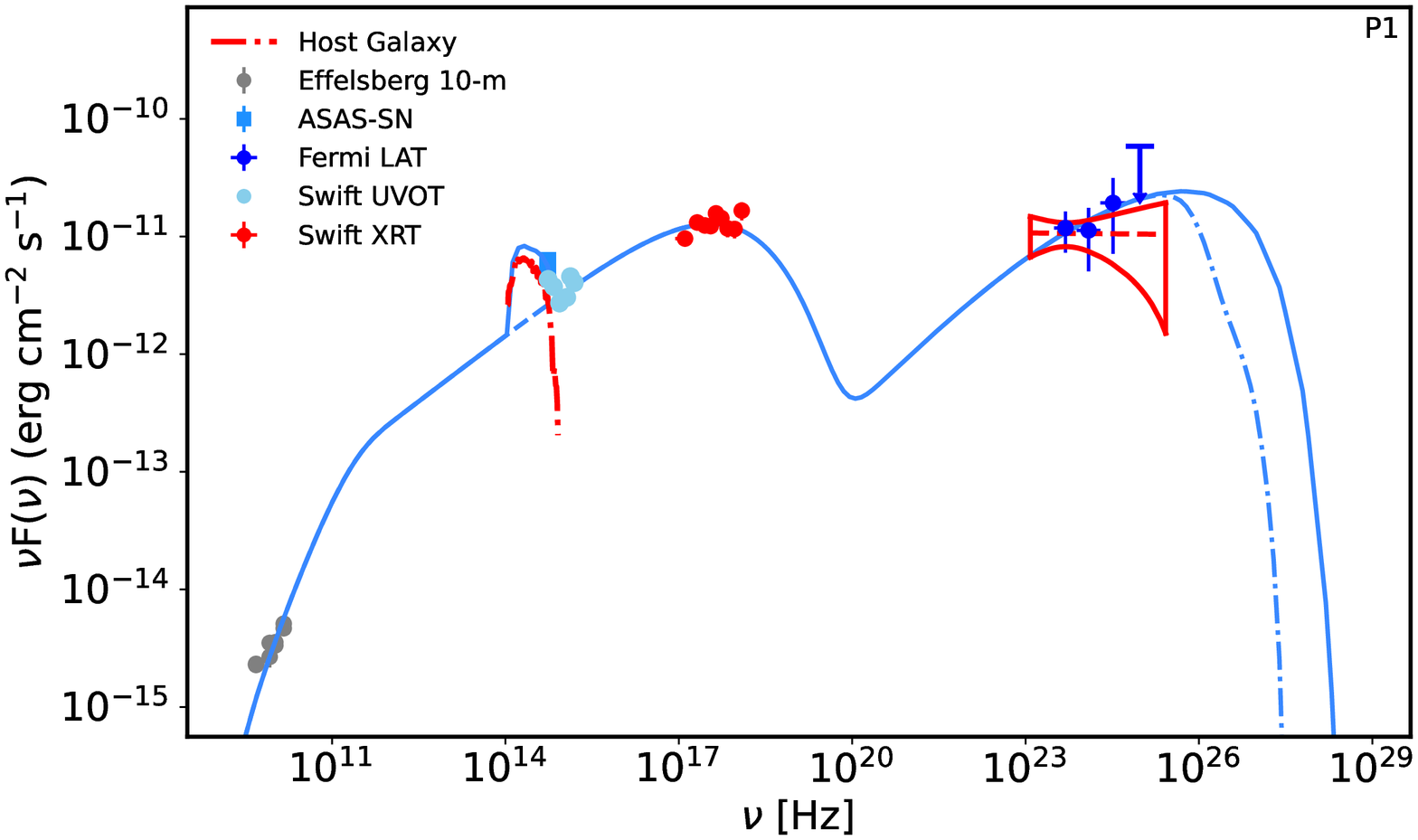}
   \includegraphics[width=0.49 \textwidth]{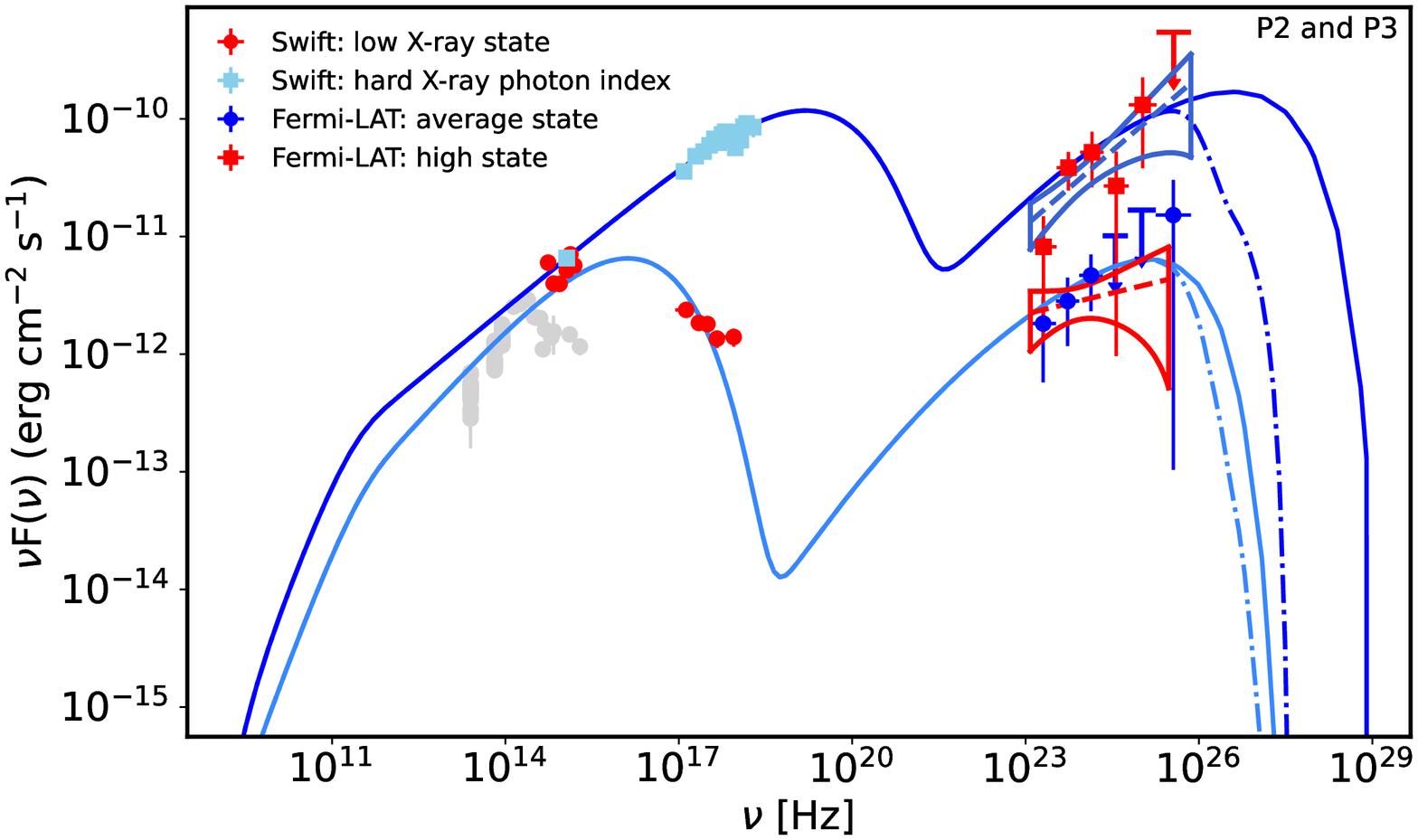}\\
      \caption{Modeling of multiwavelength SEDs of \fgl\ during the MJD $57938.35\pm12.80$ (left panel) and MJD $57989.16\pm5.16$ and MJD $58571.12\pm52.53$ (right panel; squares and circles, respectively). The model is presented in the text and the parameters are given in Table \ref{sedresults}.}
         \label{sed}
\end{figure*}
\begin{table}
\centering
\caption{Parameters of modeling of the broadband SED of \fgl\ within a simple one zone synchrotron/SSC scenario.}
\label{sedresults}	
\begin{tabular}{lccc}
\hline
Parameter		   		&P1      	&P2 & P3  \\
\hline
$\delta$       &25.6 &25.6 & 25.6     \\
$\alpha$		&2.25 & 2.19 & 1.90	\\
$\gamma^{\prime}_{\rm min}$ & 481.1 & 481.1 & 481.1\\
$\gamma^{\prime}_{\rm cut}\times$10$^{6}$ &1.16 &6.82 & 0.11 \\
B [G] $10^{-3}$             &7.82 &8.54 &13.71\\
L$_e$ [$\times10^{44}$ erg~cm$^{-3}$]           &1.60 & 2.47 & 0.50\\
L$_B$ [$\times10^{40}$ erg ~cm$^{-3}$]		&5.73 &6.83 & 17.61\\
\hline
\end{tabular}
\end{table}
The broadband SED of \fgl\, shows the usual double humped shape that can be interpreted within a common one-zone synchrotron/SSC model (Fig. \ref{sed}). It is assumed that the broadband emission is produced by relativistic electrons in a spherical blob of radius $R$ with a uniform magnetic field $B$. The emitting region is seen in the direction close to the line of sight which moves with the bulk Lorentz factor of $\Gamma\approx\delta$. It is assumed that the emitting electrons have commonly used powerlaw with an exponential cut-off energy distribution in the form of $N^{\prime}_{\rm e}(\gamma^{\prime}_{\rm e})= N^{\prime}_{0}\: \gamma^{\prime -\alpha}_{e}\:Exp[-\gamma^{\prime}_{\rm e}/\gamma^{\prime}_{\rm cut}]$ for $\gamma^{\prime}_{\rm e}\geq \gamma^{\prime}_{\rm min}$ where $\gamma^{\prime}_{\rm min}=E^{\prime}_{e}/m_{e}\:c^2$ is the minimum electron energy. These electrons interacting with the magnetic field radiate via the synchrotron mechanism which explains the low energy component in the SED while the HE hump is described by up-scattering of the synchrotron photons by the same electrons.\\
The model includes seven free parameters to describe the emitting electrons ($\alpha$, $N^{\prime}_{0}$, $\gamma^{\prime}_{\rm min}$ and $\gamma^{\prime}_{\rm cut}$) and specify the emitting region ($\delta$, $B$ and $R$). In order to reduce the number of free parameters, we assume the emission is produced in a compact region with $R=5\times10^{16}\:{\rm cm}$ which corresponds to flux changes on days time scales, as it has been found in the \gray\ band (adaptive bins in Fig. \ref{lightcurve}). In order to constrain $\delta$ and $\gamma^{\prime}_{\rm min}$ initially the SED on MJD $57938.35\pm12.8$ (P1) with (quasi) simultaneous data ranging from radio to HE \gray\ bands is modeled, resulting in $\delta=25.6$ and $\gamma^{\prime}_{\rm min}=481.1$ which are typical values usually obtained for BL Lacs. These were fixed parameters when modeling the SEDs observed on MJD 57989.16 (P2) and MJD 58571.12 (P3) for which the low energy data are lacking. In order to account for the contribution of the host galaxy (see the excess in the IR/optical part of the SED on MJD $57938.35\pm12.8$), in addition to synchrotron/SSC component also we adopted a template of giant elliptical galaxy. The fitting is performed with the open source package {\it JetSet} \citep{2006A&A...448..861M, 2011ApJ...739...66T, 2009A&A...501..879T}.\\
The SED modeling results are shown in Fig. \ref{sed} and the corresponding parameters presented in Table \ref{sedresults}. In the left panel of Fig. \ref{sed}, the Swift UVOT and ASAS-SN data \citep[downloaded from the ASAS-SN Sky Patrol web site][]{asasn}\footnote{https://asas-sn.osu.edu/)}
%(\textcolor{green}{we did not mention how we got the ASAS-SN data, maybe we should add a sentence about that ?}) 
allow to estimate or at least constrain the contribution of the host galaxy; the peak of the host galaxy component is at $6.13\times10^{-12}\:{\rm erg\:cm^{-2}\:s^{-1}\:}$. When the SED in the quiescent state is considered (P3), the data can be reproduced when $\alpha=1.96$ and $\gamma^{\prime}_{\rm cut}=1.13\times10^5$ and the magnetic field is $B=1.37\times10^{-2}\:{\rm G}$. In this case low energy component peak is at $\sim 2.42\times10^{16}\:{\rm Hz}$ which, however, shifts to higher frequencies when the active states of the source are considered; i.e., during MJD $57938.35 \pm 12.8$ the X-ray data defines the peak to be at $\sim10^{17}\:{\rm Hz}$ whereas during the hardest X-ray state the peak is shifted well beyond $\sim10^{18}\:{\rm Hz}$. This shift changes also the maximum energy up to which the particles are effectively accelerated; $\gamma^{\prime}_{\rm cut}$ increases to $1.16\times10^6$ and $6.82\times10^6$ when the SEDs on P1 and P2 are considered, respectively. The modeling resulted in a similar powerlaw index and magnetic field for these active X-ray periods, namely $\alpha=2.25$ and $7.82 \times10^{-3}\:{\rm G}$ for P1 and $\alpha=2.19$ and $8.54 \times10^{-3}\:{\rm G}$ for P2. In all the considered periods the jet is particle dominated with $U_{\rm e}/U_{\rm B}>3000$ for P1 and P2 and $U_{\rm e}/U_{\rm B}\simeq350$ for P3. However, $U_{\rm e}$ strongly depends on $\gamma^{\prime}_{\rm min}$ which is not well constrained, and when allowing $\gamma^{\prime}_{\rm min}$ to vary $U_{\rm e}/U_{\rm B}$ ratio can change. The jet power in the form of magnetic field and electron kinetic energy are calculated by $L_{B}=\pi c R^2 \Gamma^2 U_{B}$ and $L_{e}=\pi c R^2 \Gamma^2 U_{e}$,  respectively, and are given in Table \ref{sedresults}.\\
\section{Discussion and Conclusion}\label{diss}
\begin{figure}
%   \centering
   \includegraphics[width=0.49 \textwidth]{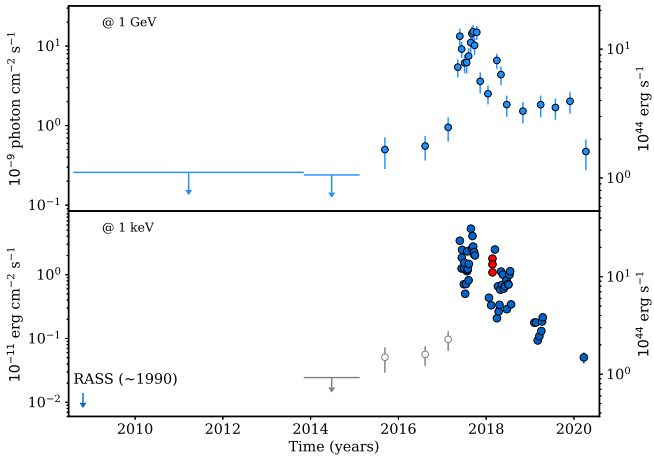}
      \caption{The \gray\ (@ 1 GeV) and the reconstructed X-ray (@ 1 keV) light curve of \fgl\ (see text for details).} 
      %In the bottom panel, the X-ray flux estimated in each orbit is shown in dark red and that of a single observation in blue. 
%      }
         \label{LC1GeV1KeV}
\end{figure}

\begin{figure*}
%   \centering
   \includegraphics[width=1.0 \textwidth]{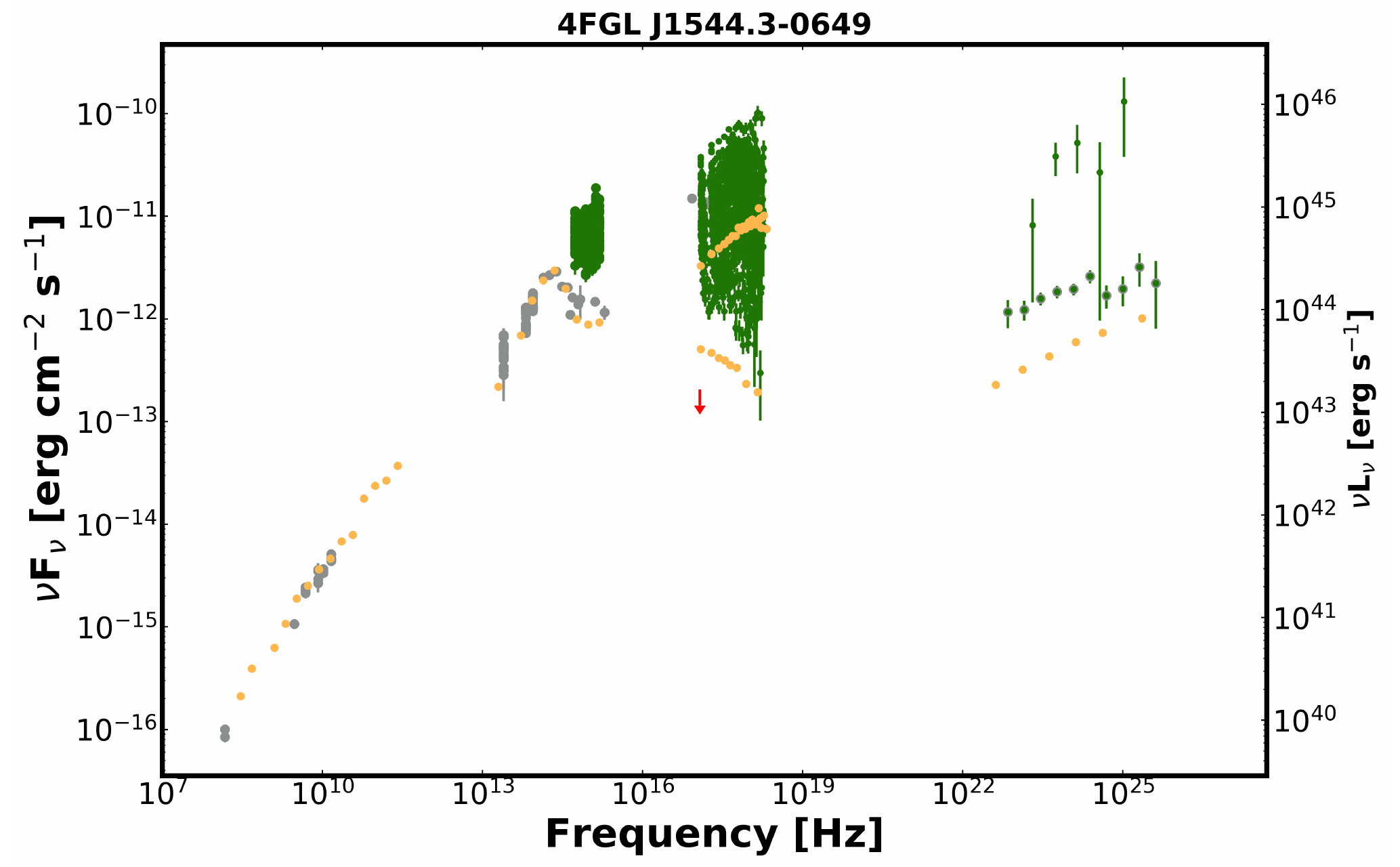}
      \caption{The SED of \fgl. Green points are from Swift UVOT, Swift XRT and Fermi-LAT; grey points are archival multi-frequency data retrieved with VOU-Blazars. For comparison the average SED of Mrk 501, scaled to  flux level of \fgl, in the radio/infrared bands, is plotted as orange points. In the X-ray band instead of the average spectrum we plot the the spectra observed in Mrk 501 when it was at maximum and minimum intensity.}
         \label{sed2}
\end{figure*}

We have presented the results of the analysis of optical/UV, X-ray and \gray\ observations of \fgl, a remarkable blazar showing transient-like high-energy multiwavelength emission, a behaviour that was never observed before in AGNs. This source, in fact, remained below the sensitivity limits of all the high-energy surveys and observations until May 2017, when it brightened to such a level to be detectable in short exposures by Fermi-LAT and by the MAXI X-ray sky monitor \citep[][]{MAXI}.\\
%\textcolor{green}{ what do you think if we will remove this paragraph ?\\
%\fgl, entered an active \gray emission phase during MJD 57869.3-58052.0 with the highest \gray flux of $(2.35\pm	0.55)\times10^{-8}\:{\rm photon\:cm^{-2}\:s^{-1}}$ above $801.6$ MeV. During this period the flux increased in the X-ray and optical/UV bands as well (though not simultaneously in the latter band) and the source behaved as a regular BL Lac. However, after the flux decreased in all the bands around MJD 58500, the source became unobservable in all the considered bands. \textcolor{red}{Is this really true? There are some detections after MJD 58500..}
%The time evolution of broadband SED is available here \href{https://www.youtube.com/watch?v=9feaNWi1RDs}{\nolinkurl{youtu.be/9feaNWi1RDs}}.}\\
During the brightening, the 2-10 keV X-ray flux strongly varies increasing from $(1.0-5.0)\times10^{-12}\:{\rm erg\:cm^{-2}\:s^{-1}}$ to $(1.28\pm0.05)\times10^{-10}\:{\rm erg\:cm^{-2}\:s^{-1}}$. %\textcolor{red}{Which band?} 
%in the bright X-ray states. 
This variability, which is even larger when the flux is estimated in each Swift orbit, is clearly associated to a harder-when-brighter behavior, as also noted by \cite{2019A&A...622A.116U} based on XMM observations. Such flux and photon index correlation is well known in HBL blazars \citep[e.g.][]{1990ApJ...356..432G}, and is usually interpreted by the relation between the acceleration and cooling processes. As shown in \citet{1998A&A...333..452K}, when the injection decreases and particles are loosing energy by synchrotron cooling or any other cooling process that is faster at higher energies, the spectrum becomes steeper.\\
During the \gray\ brightening of the source, the X-ray spectrum modifies as well. The available optical/UV and X-ray data defines the peak of the low energy component to be within $\nu_{\rm peak-s}=10^{15}-10^{17}$ Hz, implying \fgl\ is a classical HBL. However, occasionally (e.g., on MJD 57898.70, 57989.16, 57998.92, 58009.40 and 58021.40, see the SED/Light curve animation) the source shows a synchrotron component peak above $10^{18}$ Hz which is more typical for extreme synchrotron BL Lacs. Yet, in all these periods the \gray\ photon index hardens as well, e.g., $1.74\pm0.20$ on MJD 57898.70, $1.57\pm0.18$ on MJD 57989.16 or $1.61\pm0.19$ on MJD 58021.40, shifting the HE peak to VHEs. The peaks in the SED are determined by $\gamma^{\prime}_{\rm cut}\simeq \sqrt{\frac{\nu_{\rm c,\:peak}}{\nu_{\rm s,\:peak}}}$ implying in P1 and P2 when $\nu_{\rm peak}\geq10^{18}$ Hz the second peak is around $2\times10^{26}-2\times10^{27}$ Hz ($1-10$ TeV) (solid line in Fig. \ref{sed}). However, this peak and the flux is suppressed by Klein Nishina effects and EBL absorption, which for the distance of \fgl\ starts at $\sim 5\times10^{25}$ Hz ($\sim200$ GeV; dot dashed line in Fig. \ref{sed}). Despite this reduction, during the bright \gray\ and X-ray states the source flux at 1 TeV is expected to be  $\simeq10^{-11}\:{\rm erg\:cm^{-2}\:s^{-1}}$, well above the sensitivity of the current ground-based Cherenkov telescopes. In addition, the hard \gray\ photon index observed in the MeV/GeV band and the applied SSC model predict also a very hard photon index in TeV band after EBL correction. So, \fgl\ mimics properties similar to those of extreme BL Lacs in \grays\ \citep{2015MNRAS.451..611B,2011MNRAS.414.3566T}, which are an interesting sub-class not only because of their possible association with VHE neutrinos and cosmic rays \citep{2016MNRAS.457.3582P, 2017MNRAS.468..597R} but also because they pose major difficulties for the understanding of particle acceleration and emission processes in blazar jets.\\
The multiwavelength light curve of Fig. \ref{lightcurve}, shows that the optical/UV flux increase is delayed compared to that in the X-ray and \gray\ bands. In general, lags in different energy bands are expected either from changes in the global parameters (magnetic field, Lorentz factor, etc.) or from the temporal evolution of emitting particles. For example, \citet{2008A&A...486..679B} proposed that if the low energy component is produced when the blob is closer to the base of the jet and the HE photons are emitted at larger distances from the jet base, at which the blob has accelerated to a higher Lorentz factor, this could result in delays between the low energy and HE flares. Or, in complex jet geometry if the flare is caused by the collision between a relativistic shock wave and a stationary feature in the jet, it results in a complex variation of fluxes in different bands \citep{2004ApJ...613..725S}. However, a simple consideration of particle acceleration and cooling can naturally explain the observed delay in the multiwavelength light curve of \fgl. The SED in different epochs is well reproduced by a one-zone synchrotron/SSC model with physically reasonable parameters. Within this model, the flare and modification of the spectra can be interpreted by an increase in the luminosity of electrons and $\gamma^{\prime}_{\rm cut}$; during the bright X-ray period the luminosity increased $\sim4.9$ times whereas $\gamma^{\prime}_{\rm cut}$ $\sim62$ times. This implies that the X-ray and \gray\ flare was most likely caused by injection of new freshly accelerated electrons. The cooling time of these electrons is 
\begin{equation}
t_{\rm cooling}=\frac{3}{4}\:\frac{m_{\rm e}\:c}{\sigma_{\rm T}(u_{\rm B}+u_{\rm syn})\:\gamma}=\frac{6\:\pi\:m_{\rm e}\:c}{\sigma_{\rm T}\:\gamma\:B^2\:(1+u_{\rm syn}/u_{\rm B})}.   
\end{equation}
where $u_{\rm syn}$ is the synchrotron photon energy density in the jet frame. Considering the peak frequency of synchrotron emission is $\nu=1.2\times10^6\:B\:\delta\:\gamma^2$, the cooling time of the electron emitting photons with frequency $\nu_{15}=\nu/10^{15}\:{\rm Hz}$ is $t_{\rm cooling}=2.7\times10^4\:(\delta/1+z)^{0.5}\:B^{-1.5}\:\nu_{15}^{-0.5}$ s. Considering the parameters obtained for the P1, the delay between X-ray emission ($10^{18}$ Hz) and optical/UV ($5\times10^{14}$ Hz) is of the order of $\sim114$ days which is very similar to that observed. However, in order to investigate in details the time lags and thus to infer the underlying physics, much denser monitoring of \fgl\ in optical/UV band is needed.\\
Figure \ref{LC1GeV1KeV} plots the high-energy light-curves of \fgl\,  spanning the entire operational period of the Fermi mission. The upper panel, reporting the \gray\ data, illustrate the transient nature of this source, which was not detected by the LAT instrument until late 2015, peaked in 2017-2018 and then faded.
The lower panel reconstructs the long-term light-curve of \fgl\, in the X-ray band, combining the Swift-XRT measurements (blue points), with the XMM data from \cite{2019A&A...622A.116U} (red points) and the expected X-ray flux level estimated from the flux in the \gray\ band, based on the average x-ray/\gray\ flux ratio (open grey circles). An upper limit derived from the non-detection of the source in the RASS X-ray sky survey, which was carried out in 1990-1991 \citep{Voges1999}, is also shown in the leftmost part of the plot. The reconstructed X-ray light-curve implies a rise time of approximately two years, and a decay of approximately three years.  

The SED of \fgl\ (Fig. \ref{sed2}) assembled with Swift-XRT, Swift-UVOT, Fermi-LAT (green points) and archival data (grey points), shows a highly variable X-ray spectrum, changing its intensity by large factors and often peaking in the X-ray band. For comparison the average SED of the well known HBL blazar Mrk501, rescaled to match the radio emission of \fgl, together with its X-ray spectra observed during its minimum and maximum intensity state, is plotted as light orange points in the same figure. The nearly identical shape of the radio to optical SEDs and the range of observed X-ray fluxes, confirms that \fgl\ during the active phase is a typical HBL blazar, with X-ray and \gray\ emission that are even stronger than that of Mrk 501. The upper limit from the RASS all sky survey (red arrow) sets the magnitude of the intensity variations in the soft X-ray band, which ranged from below $\approx 10^{-13}\:{\rm erg\:cm^{-2}\:s^{-1}}$ in 1990 to such a high value to place \fgl\ amongst the 20 brightest X-ray blazars.

The existence of transient (HBL) blazars was never considered in the literature before the discovery of \fgl\ and, consequently, never taken into account in population studies. However, if this dual behaviour 
is relatively common, it might have implications in blazar research, especially in high-energy and multi-messenger astrophysics, depending on how numerous transient blazars are and how frequently they enter a bright phase.  
We don't know if \fgl\ represents the tip of the iceberg, or it is simply an isolated event. 
The discovery of \fgl\ is a clear selection effect.
%since this blazar flared to be one of the brightest hard \gray\ sources in the sky, reaching the threshold for a detection in a few days exposures. 
Similar transient sources would probably go unnoticed in case their top flux was only a factor of a few lower than that of \fgl, or with a softer \gray\ spectrum, or even in cases where the \gray\ intensity remains low during strong X-ray flares associated to neutrino emission as in scenarios similar to that presented by \cite{MP}.
The identification of similar objects would add crucial information to our knowledge of the population of extragalactic sources, especially in the high-energy domain.
Areas of research that might be affected would be e.g. the determination of the LogN-LogS, luminosity function and cosmological evolution of HBL blazars, the identification of still unassociated \gray\ and VHE sources, and the contribution of discrete sources to the \gray\ and possibly high-energy neutrino cosmic backgrounds. A transient nature of some electromagnetic counterparts of astrophysical neutrinos would on one side increase the level of difficulty in the process of the association, but on the other it would strengthen the association in case X-ray or \gray\ flares are detected in association with the neutrino arrival. 
%It is not possible to quantify the magnitude of this contribution based on a single source and episode, but the strength of \fgl , which for about one year was so bright to be among the 1\% brightest blazars in the X-ray sky, allow us to consider the case in which fainter episodes of this type in many blazars might have gone unnoticed. 
A recent study \citep[][]{GiommiTracks} reported a 3.23 $\sigma$ post trial excess of HBL Blazars in the error regions of a sample of 70 Ice Cube astrophysical neutrinos with good positional accuracy. If HBL blazars are indeed counterparts of high-energy neutrinos, some of them could be transient sources somewhat fainter than \fgl\ that would difficult to locate since they are not listed in catalogues and no full coverage of the still large Ice Cube neutrino positional uncertainty can be achieved with current X-ray imaging instruments.      
A full census of these so far elusive objects would be possible only when sensitive all-sky X-ray (or \gray) monitor detectors become available.
Meanwhile a careful search for weak transients in the Fermi-LAT data on the positions of flat-spectrum radio sources with radio to infrared flux ratio similar to that of HBL objects could lead to the discovery of additional transient blazars and set limits to their duty cycle and space density.

\section*{Acknowledgements}

We acknowledge the use of data, analysis tools and services from the Open Universe platform, the ASI Space Science Data Center (SSDC), the Astrophysics Science Archive Research Center (HEASARC), the Fermi Science Tools, the All-Sky Automated Survey for Supernovae (ASAS-SN), the Astrophysics Data System (ADS), and the National Extra-galactic Database (NED).\\
\textbf{PG} acknowledges the support of the Technische Universit\"at M\"unchen - Institute for Advanced Studies, funded by the German Excellence Initiative (and the European Union Seventh Framework Programme under grant agreement no. 291763) and support by the DFG Cluster of Excellence "Origin and Structure of the Universe".\\
\textbf{NS} acknowledges the support by the Science Committee of RA, in the frames of the research project No 20TTCG-1C015.\\
This work used resources from the EGI infrastructure with the dedicated support of CESGA (Spain).\\
We are grateful to M. Petropoulou, P. Padovani and R. Middei for useful comments.

\section*{Data availability}
All the Swift-XRT spectra, best fit values and other imaging analysis results will be made available as part of a publication presenting the analysis of XRT data of all the blazars frequently observed by Swift (Giommi et. al. 2020, in preparation), in the framework of the Open Universe for blazars program. 
The remaining data underlying this article will be shared on reasonable request to the corresponding author.
%%%%%%%%%%%%%%%%%%%%%%%%%%%%%%%%%%%%%%%%%%%%%%%%%%



\bibliographystyle{mnras}
\bibliography{biblio} % if your bibtex file is called example.bib

%%%%%%%%%%%%%%%%%%%%%%%%%%%%%%%%%%%%%%%%%%%%%%%%%%

% Don't change these lines
\bsp	% typesetting comment
\label{lastpage}
\end{document}